\documentclass[11pt]{article}
\usepackage{jheppubmod}
\pdfoutput=1
\usepackage{framed}
\usepackage{psfrag}
\usepackage{array}
\usepackage{bm, bbm, bbold}
\usepackage{amssymb}
\usepackage{amsmath}
\usepackage{amsthm}
\usepackage{braket}
\usepackage{hyperref}
\usepackage{graphicx}
\usepackage[labelsep=quad]{subcaption}
\usepackage{color,soul}
\usepackage[punctsep]{collref}
\collectsep[]{;}
\usepackage{epsfig}
\usepackage{wrapfig}
\usepackage{tikz}
\usetikzlibrary{arrows,plotmarks,positioning,calc}
\usepackage{environ}
\NewEnviron{eqn}{
\begin{align}
\begin{split}
  \BODY
\end{split}
\end{align}
}
\NewEnviron{eqn*}{
\begin{align*}
\begin{split}
  \BODY
\end{split}
\end{align*}
}
\newcommand{\be}{\begin{equation}}
\newcommand{\ee}{\end{equation}}
\newcommand{\bes}{\begin{equation*}}
\newcommand{\ees}{\end{equation*}}

\usepackage[
    natbib=true,
    style=numeric,
    sorting=none, 
    maxbibnames=99
]{biblatex}
\def \scrip{{\cal I}^{+}}
\def \scrim{{\cal I}^{-}}
\def\e{\epsilon}
\def\q{\theta}

\def\p{\partial}
\def\O{\Omega}

\newcommand{\intinf}{\int_{-\infty}^{\infty}} 
\newcommand{\intsinf}{\int_{0}^{\infty}} 
\newcommand{\dagg}{\dagger}

\newcommand{\vac}{\ket{\{s\}}}

\begin{document}    
\title{Recovering information in an asymptotically flat spacetime in quantum gravity}
\author[a]{Chandramouli Chowdhury}
\author[b]{and Olga Papadoulaki}
\affiliation[a]{International Centre for Theoretical Sciences, Tata Institute of Fundamental Research, Shivakote, Bengaluru 560089, India.}
\affiliation[b]{Perimeter Institute for Theoretical Physics, Waterloo, ON N2L 2Y5, Canada}
\emailAdd{chandramouli.chowdhury@icts.res.in}
\emailAdd{opapadoulaki@perimeterinstitute.ca}

\abstract{As an extension of arXiv:\{2002.02448, 2008.10740\} we present a physical protocol that a set of observers can use to detect a pure state in the bulk when they are spread across a small cut near $\scrip_-$ in flat spacetime.  The protocol involves the modification of a bulk state using simple unitary operators and measurements of the energy of the state. The states that we study are constructed by acting with low energy operators on a vacuum state such that a perturbative analysis is valid. We restrict ourselves to $3+1$ dimensional spacetimes and only consider massless excitations. From this analysis, the principle of holography of information becomes manifest in the case of asymptotically flat spacetime.}

\maketitle
\noindent
\flushbottom

\section{Introduction and Setup}
In \cite{Chowdhury:2020hse} it was shown that a set of observers living on a thin time band in global Anti-de Sitter space (AdS) can determine a state in the bulk via a physical protocol in a theory of quantum gravity. The existence of such a protocol is a consequence of the {\it principle of holography of information}, which was established in a series of papers \cite{Laddha:2020kvp, Raju:2019qjq, Chowdhury:2021nxw} and is reviewed in \cite{Raju:2020smc}. For a theory of quantum gravity in flat spacetime, this states that all information about a state in the bulk is also available at a small cut near the past of the future null infinity $\scrip_-$ (or near the future of past null infinity $\scrim_+$). The validity of the principle relies on the following basic properties of semi-classical gravity as elucidated by DeWitt \cite{DeWitt:1967yk} and later expanded in \cite{Laddha:2020kvp, Raju:2019qjq, Chowdhury:2021nxw}:
\begin{enumerate}
 \item Any finite energy excitation in the bulk must leave an imprint at the boundary due to the uncertainty principle. As a corollary, this also implies that there are no local gauge invariant operators in gravity. 
 
 \item The Hamiltonian of the theory is a boundary term\footnote{The bulk Hamiltonian is set to zero by the constraints in phase space.} and can be expressed in terms of the metric fluctuations at the boundary (eg: see equation \eqref{energy}). 
 
 \item We assume that the Hamiltonian \eqref{energy} remains positive in the quantum theory as it is quadratic in the degrees of freedom.
 
 \item The eigenstate of the Hamiltonian with the least energy is defined to be the vacuum state and its energy is renormalized to zero.

\end{enumerate}
We will require these properties to hold in perturbative gravity and will be using them throughout the paper.

The best known example of the principle of holography of information is the AdS/CFT correspondence \cite{adscft, Witten:1998qj}. However the authors of \cite{Laddha:2020kvp} demonstrated how the principle works in $3+1$ dimensional flat spacetime with gravity coupled to massless matter. They showed that all information on a Cauchy slice in the bulk is also available at $\scrip_-$ (or $\scrim_+$). Motivated by this, we follow a similar program as in \cite{Chowdhury:2020hse} and establish a physical protocol for detecting a certain class of excitations in $3+1$ dimensional flat spacetime in a theory of gravity coupled to massless matter fields.
\begin{figure}[h]
\begin{center}
\scalebox{0.5}{\includegraphics{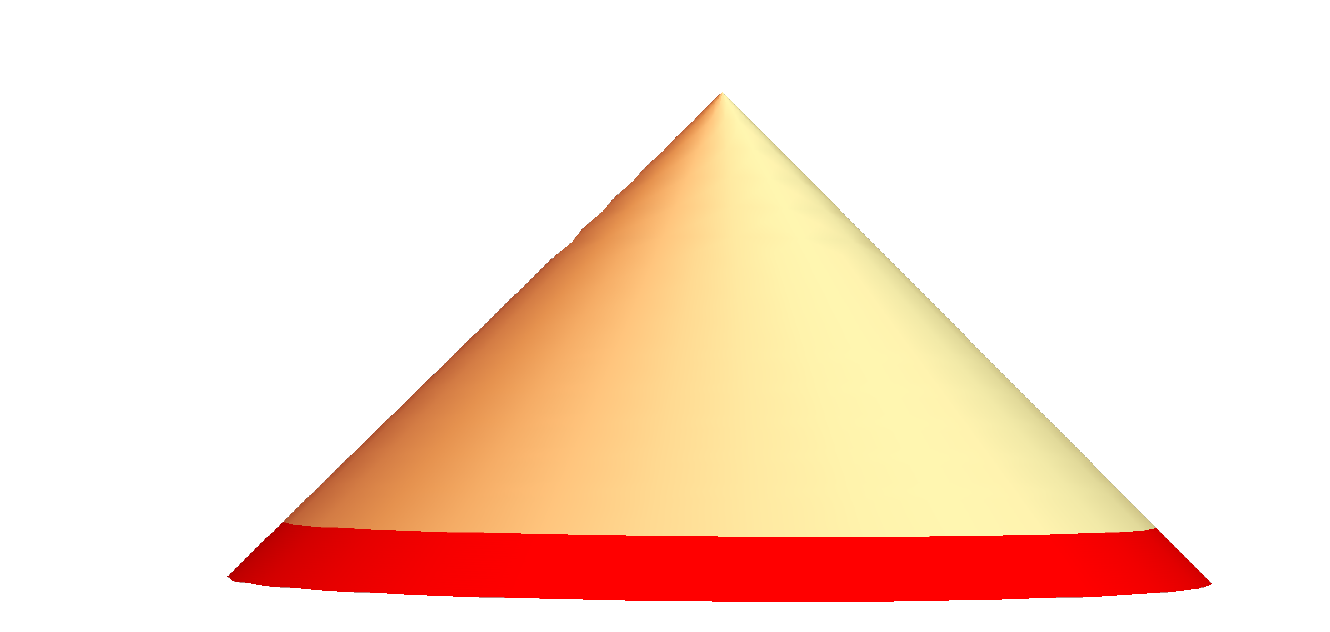}} 
\begin{tikzpicture}
\node at (0, 0) {$u= -\infty$};
\node at (-0, 0.5) {$u= -1/\epsilon$};
\end{tikzpicture}

 \caption{Location of observers near $\scrip_-$ in Flat spacetime. The state $\ket\psi$ is composed of operators spread across all of $\scrip$ (denoted by the yellow + red region of the cone) but the measurements are only performed in the red region. }
 \label{fig1}
 \end{center}
\end{figure}
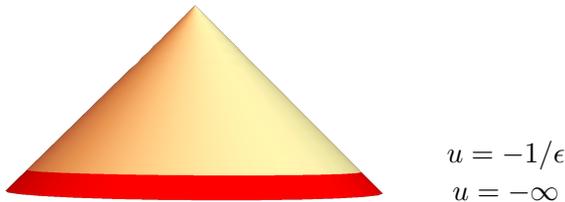

The observers are localized in a small region near $\scrip_-$ as depicted by the red band in figure(\ref{fig1})\footnote{There are analogous statements about $\scrim$ (and the localization of information at $\scrim_+$), however for concreteness, we shall work only from the point of view of $\scrip$.}. They are given the task of determining a state $\ket{\psi}$ by performing certain measurements. These include the measurement of conserved charges (localized at $\scrip_-$) and also a modification of the state $\ket{\psi}$ by unitary operators near $\scrip_-$. The latter restriction is necessary because in any physical process the observers can modify the interaction Hamiltonian (between themselves and the system) by Hermitian operators. In perturbation theory, this results in the action of a unitary operator acting on the state. 

It has been shown \cite{Laddha:2020kvp} that the knowledge of all possible correlation functions composed of operators near $\scrip_-$, in a particular state, allows one to reconstruct the state itself. However in this paper we are restricting ourselves to measurements which can be performed by observers localized at $\scrip_-$ in a ``physical experiment". For this purpose we shall only consider a class of states $\ket\psi$ which are built by exciting the vacuum state $\ket{0}$ by low energy operators. There are more general classes of states as discussed in \cite{Marolf:2015jha} but the examination of these and a construction of a protocol to detect them from $\scrip_-$ are left for future work.

We now discuss the difficulties in constructing such a protocol in flat spacetime as compared to AdS. Flat spacetime has an intricate IR structure due to the presence of an infinite dimensional symmetry \cite{He:2014laa, Strominger:2014pwa, Strominger:2017zoo}  which leads to an infinitely  degenerate vacuum state. 
Thus  flat spacetime has multiple vacua in contrast to the AdS spacetime that has a unique vacuum.

Another major difficulty in flat spacetime is the lack of a discrete energy spectrum (as is the case for the AdS spacetime). Since the energy is continuous in flat spacetime it is not useful to decompose a state $\ket{\psi}$ in terms of a basis spanned by energy eigenstates. However, we can still exploit the fact that there is a lower bound of the energy when measured from asymptotic infinity, which we renormalize to zero. Subsequently, we introduce a more convenient basis to decompose the states. This is called the {\it normal-ordered basis}, which is very similar to the usual Fock basis (see eq.  \eqref{psidefn1}). Throughout this paper we represent the matter fields with scalars but this restriction can be relaxed and we can consider fields of any spin (or even stringy excitations). By causality, any massless excitation can be expressed in terms of operators smeared over $\scrip$ and therefore the basis is constructed out of fields living at $\scrip$. A similar basis construction is also possible in AdS spacetime but in that case it was more convenient to work with an energy eigen basis \cite{Chowdhury:2020hse}. 

In this paper we shall restrict to measuring the energy of states built using {\it hard operators} (an operator with finite energy) on a particular vacuum state denoted by $\ket{0}$. Using the Born-rule, this  reduces to computing expectation values of a product of unitary operators and the projector onto the vacuum state. We will show that this allows us to completely decode a state of the kind shown in eq.\eqref{psidefn1}. 

As discussed below, such a protocol does not violate causality. This is because, in order to measure the energy of a state, the observers are forced to be spread over the whole Celestial sphere since the energy is expressed  as a surface integral of metric fluctuations over the entire Celestial sphere (see equation \eqref{energy}). Hence each observer  only measures a part of the metric fluctuation and in order to evaluate the energy, they have to meet and sum their results. This process clearly takes more than the light crossing time and therefore prevents any violation of causality.

We end this section by stating why such a protocol only works in a theory of gravity and not in other gauge theories or Local QFTs (LQFT). In an ordinary LQFT (without dynamical gravity) such a protocol would clearly fail since there exists operators which commute with {\it all} operators at $\scrip_-$ because of micro-causality.  Therefore it is not possible to distinguish a state $\ket\psi$ from the state $U_{bulk} \ket{\psi}$ where $U_{bulk}$ is any unitary operator in the bulk. Similar local operators also exist in a gauge theory (example: in QED we have $F_{\mu\nu}^2$) which prevent the reconstruction of information in those theories. This counter argument does not work in quantum gravity as the notion of micro-causality relies on {\it spacelike separation} and since the metric itself fluctuates at $O(\sqrt{G_N})$ \cite{Arkani-Hamed:2007ryv, Donnelly:2015hta} in a theory of quantum gravity, micro-causality is violated perturbatively in $O(\sqrt{G_N})$. Even though this is a tiny violation, it can be observed in the regime of effective field theory and this will become apparent in the main text.

We review the Hilbert space structure of flat spacetime in section \ref{sec:hilbert}, then properly define the main question and its solution in section \ref{sec:main} and end with conclusions and discussions on various related ideas in section \ref{discussion}.

\section{Hilbert Space of Flat Spacetime}\label{sec:hilbert}
In this section we review the asymptotic structure of Minkowski spacetime and how its information is encoded in data at null infinity \cite{Laddha:2020kvp, Ashtekar:1981sf, Ashtekar:1987tt}. Moreover, we will revisit how upon quantization these data give rise to the Hilbert space of the low energy effective theory. 

The metric for an asymptotically flat space time near $\scrip$ can be written in the retarded Bondi coordinates \cite{Bondi:1962px, Sachs:1962wk} as
    
\be\label{metric}
ds^2 = - du^2 - 2 du dr + r^2 \gamma_{AB} d\Omega^Ad\Omega^B + r C_{AB} d\Omega^{A}d\Omega^{B} + \frac{2m_{B}}{r} du^2 +\gamma^{AC}D_{C}C_{AB} dud\Omega^B +...\,,
\ee
where the {\it retarded time} $u =t- r$ with $t$ being the time, $r$ the radial direction and $\Omega =(\theta,\phi)$ the coordinates on the unit sphere $S^{2}$. The capital latin letters $A,\cdots$ take values on the unit sphere and $\gamma_{AB}$ is the metric of the unit $S^{2}$. $C_{AB}(u, \Omega)$ is the shear field and it encodes the information about the radiative degrees of freedom. In this gauge $\gamma^{AB}C_{AB}=0$, thus the shear is traceless. The radiative data are encoded at future null infinity $\mathcal{I}^{+}$ which has the topology $\mathbb R \times S^{2}$ and is parametrized by $(u,\Omega)$. The afforementioned unit $S^2$ is called the {\it Celestial sphere}. 
$D_{A}$ is the covariant derivative with respect to the metric $\gamma_{AB}$ and $m_{B}$ is the {\it Bondi mass aspect}.

Throughout this paper we shall work with matter fields which are massless scalar fields, however the results are easily generalizable to other massless fields. The large$-r$ fall-off  of the scalar field is fixed by demanding the finiteness of its energy 
\be\label{falloff}
\lim_{r \to \infty}\phi^{bulk}(u, r,\Omega) = \frac{1}{r}\phi (u,\Omega) + \mathcal{O}\Big(\frac{1}{r^2}\Big)\,,
\ee
where $\phi(u, \Omega)$ encodes the classical radiative data of the matter field.

At null infinity the gravitational and matter data are not independent as they are related by the Hamiltonian constraint of General relativity. This can be explicitly seen from the $uu$-component of the Einstein equation which gives the evolution equation for the Bondi mass aspect

\be\label{ratemb}
\partial_{u}m_{B}=\frac{1}{4} D^{A}D^{B} N_{AB}-\frac{1}{8}N_{AB}N^{AB}-4\pi G T_{uu}^{(0)}\,,
\ee
where $N_{AB}=\partial_{u}C_{AB}$ is the Bondi News tensor and $T_{uu}^{(0)}$  is the leading order of the matter stress tensor,  which for a scalar field is given by $\frac{1}{2}(\partial_{u}\phi)^{2}$.  Thus $m_{B}$ is a function of the radiative data and of an integration constant at $u=-\infty$ (the Bondi mass at $\scrip_+ $ is zero since we do not consider massive particles here).

We move forward to introduce the phase space and the conserved charges. The radiative phase space is characterized in terms of the News and the Shear tensor. The Poisson brakets between the two was found to be \cite{Ashtekar:1981sf, Ashtekar:1987tt, Ashtekar:1981bq}
\be
\left\lbrace N_{AB}(u,\Omega), C_{MN}(u^\prime, \Omega^{\prime})\right\rbrace= -\frac{16\pi}{\sqrt{\gamma}}\delta(u-u^{\prime})\delta^{2}(\Omega-\Omega^{\prime})\left[\gamma_{A(M}\gamma_{N)B}-\frac{1}{2}\gamma_{AB}\gamma_{MN}\right]\,.
\ee
This will be later promoted to a commutator upon quantizataion. 

In the absence of massive particles, the conserved charges of the theory can be expressed in terms of the integration constant of $m_B$ at $u \to -\infty$. These charges are known as the {\it supertranslation charge} 
\be\label{supert}
\mathcal{Q}_{l,m}=\frac{1}{4\pi G_N}\int\sqrt{\gamma}d^2\Omega \ Y_{l,m}(\Omega) m_{B}(u=-\infty,\Omega)
\ee
where $Y_{l,m}$ are the spherical harmonics on $S^2$. $\mathcal{Q}_{l=0,m=0}$ is the familiar ADM Hamiltonian. The supertranslation charges can be separated into a {\it soft} and a {\it hard} part by using eq.\eqref{ratemb}. This decomposition can also be thought of as a separation into terms involving linear and non-linear News 
\begin{subequations}
 \be\label{soft}
\mathcal{Q}^{soft}_{l,m} = -\frac{1}{16\pi G_N}\int_{-\infty}^{\infty}du\,d^2\Omega\,\sqrt{\gamma}Y_{lm}(\Omega)\left(D^{A}D^{B}N_{AB}\right)\,,
\ee
\be\label{hard}
\mathcal{Q}^{hard}_{l,m} = \frac{1}{16\pi G_N}\int_{-\infty}^{\infty}du\,d^2\Omega\,\sqrt{\gamma}Y_{lm}(\Omega)\left(\frac{1}{2}N_{AB}N^{AB}+16\pi G_N T^{M(0)}_{uu}\right)\,.
\ee
\end{subequations}

The action of the supertranslation generator on the radiative data at $\mathcal{I}^{+}$ is given by the following poisson brackets

\be
\left\lbrace C_{MN}(u,\Omega),\mathcal{Q}_{lm}\right\rbrace = Y_{lm}(\Omega) \partial_{u}C_{MN}(u,\Omega)-2\Big(D_{M}D_{N}Y_{lm}(\Omega) - \frac{1}{2} \gamma_{MN} D^2 Y_{lm}(\Omega)\Big)\,.
\ee

The radiative phase space of asymptotically flat spacetimes at future null infinity is given by free fields even at the non-linear level and the quantization of such a theory was developed by Ashetakar, et. al \cite{Ashtekar:1981sf, Ashtekar:1987tt, Ashtekar:1981bq}. In the quantum theory one derives the following commutation relations (obtained by promoting the Poisson brackets above to commutators)
\bes
\left[N_{AB}(u,\Omega),N_{CD}(u^{\prime},\Omega^{\prime})\right]=16\pi i G_N\frac{1}{\sqrt{\gamma}}\partial_{u}\delta(u - u') \delta^{2}(\Omega-\Omega^{\prime})\left(\gamma_{A(C}\gamma_{D)B}-\frac{1}{2}\gamma_{AB}\gamma_{CD}\right)\,,
\ees

\bes
\left[C_{AB}(u,\Omega),N_{CD}(u^{\prime},\Omega^{\prime})\right]=-8\pi i G_N \frac{1}{\sqrt{\gamma}} \text{sign}(u-u^{\prime})\delta^{2}(\Omega-\Omega^{\prime})\left(\gamma_{A(C}\gamma_{D)B}-\frac{1}{2}\gamma_{AB}\gamma_{CD}\right)\,,
\ees
where the tensors $C_{AB}$, $N_{AB}$ have been promoted to operators.

\subsection{Hilbert Space}
  The naive Hilbert space construction leads to states with divergent norms \cite{Ashtekar:1987tt}, but as we will present below (we refer the reader to section 2.3 of \cite{Laddha:2020kvp} for an extensive discussion) this can be resolved by defining the Hilbert space as a direct sum over Fock spaces. Each such Fock space is built on top of a specific vacuum defined by the soft part of the supertranslation charges\footnote{We thank Alok Laddha for explaining many issues about the IR structure of flat spacetime.}. 

The vacuum\footnote{An equivalent construction of the vacuum state is by considering the eigenstates of the shear mode. See \cite{Ashtekar:2018lor} for a detailed discussion.} is specified by the eigenvalue of the supertranslation charge with $l > 0$, i.e, the zero mode of the News
\be
\mathcal{Q}_{lm} \vac =s_{lm}\vac 
\ee
where $s_{lm} \in \mathbb R$ are also the eigenvalue of the soft part of the super-translation  charge since the hard part annihilates the vacuum. Therefore in order to completely specify a vacuum state we need to specify the value of $\{s\} \equiv (s_{00}, s_{1-1}, s_{10}, \cdots)$. This means the vacuum is infinitely degenerate with a degeneracy of ${\mathbb R}^{\mathbb Z}$.

We normalize the soft vacua by using a Dirac-delta normalization\footnote{Another convenient choice for normalizing the vacuum is to use the Kronecker delta function.}
\be\label{normalization}
 \braket{\{s\}|\{s'\}}= \delta(\{s\} - \{s'\}) \equiv \prod_{lm}\delta(s_{lm}-s^{\prime}_{lm})\,.
\ee

By acting with the creation operators on each $\vac$, we construct the Fock space $\mathcal{H}_{\left\lbrace s\right\rbrace}$. The total Hilbert space is given from the direct sum

\be\label{hilbert_sp}
\mathcal{H}= \bigoplus_{\left\lbrace s\right\rbrace}\mathcal{H}_{\left\lbrace s\right\rbrace}~.
\ee

To summarize, the Hilbert space of  massless states is given by the direct sum of the Fock spaces built using excitations on all possible vacua by acting with operators at $\mathcal{I}^{+}$. In \cite{Laddha:2020kvp} it was shown that one can reconstruct the aforementioned Hilbert space by acting on all possible vacua with operators defined in a small cut near the past of future null infinity $\mathcal{I}^{+}_{-}$. These operators form an algebra which we symbolize as $\mathcal{A}_{-\infty,\epsilon}$ and comprise the set of all functions of operators $C_{AB}(u,\Omega),\,\phi(u,\Omega),\,m_{B}(u,\Omega)$ at $\mathcal{I}^{+}$ with  $u\in (-\infty,-\frac{1}{\epsilon}]$. In our paper we explain how -- under certain assumptions -- observers with access to operators at a small cut near $\mathcal{I}_{-}^{+}$ can reconstruct states by performing specific physical measurements.

\subsection{Projector onto Vacuum state}\label{sec:vacstate}
Having defined the Hilbert space of the theory we now define the  vacuum state $\ket{0}$ of interest. Since the Hilbert space is a direct sum of the superselection sectors \eqref{hilbert_sp}, the  vacuum $\ket{0}$ can be expressed as a superposition of the soft vacua $\ket{\{s\}}$ 
\be\label{vacuum}
\ket{0} = \intinf  \Big( \prod_{l,m} ds_{l,m} \Big)  q_{\{s\}} \ket{\{s\}} \equiv \int Ds \ q_{\{s\}} \ket{\{s\}}
\ee
where the smearing functions $q_{\{s\}}$ are chosen such that $\ket{0}$ is normalizable\footnote{Using a Kronecker-delta normalization in \eqref{normalization} would allow us to choose $\ket{0}$ equal to a particular value of $\vac$ instead of smearing over all of them.}. The vacuum is normalized as $\braket{0|0} = 1$ and this constrains the smearing functions $q_{\{s\}}$
\be
\int Ds  |q_{\{s\}}|^2 = 1\,.
\ee
The explicit structure of $q_{\{s\}}$ is not required and we can chose any function which obeys the normalization above.  By definition, the state $\ket{0}$ is annihilated by the annihilation operators in the Fock space and it is also renormalized such that it has zero energy. In this paper we shall restrict to detecting states which are built by acting with hard operators on $\ket{0}$.

It will also be useful to define the projector onto states with zero energy\footnote{It is more physical to consider a projector onto a thin band of energies near zero and it can be checked that such a projector (when appropriately normalized) tends to $P_0$  when the band size is close to zero.}. Since the vacuum is the only state with zero energy in gravity, the projector onto states with zero energy is equivalent to the projector onto the vacuum state. The projector onto zero energy eigenstates can be expressed as \cite{Laddha:2020kvp}
\be\label{zeroprojector}
P_0 = \int Ds \ket{\{s\}}\bra{\{s\}}. 
\ee
Since energy is measured using the ADM Hamiltonian, this projector is an element of the algebra of operators at $\scrip_-$.

\section{Physical Protocol for detecting Massless particles in Flat spacetime}\label{sec:main}
In this section we extend the main result presented in \cite{Chowdhury:2020hse} for the case of quantum gravity coupled to massless fields in 3+1 dimensional flat spacetime\footnote{We will restrict to $3+1$ dimensional spacetime but it should be possible to generalize our results to any even dimensional spacetime.}. As described in the previous section, the vacuum in flat spacetime is infinitely degenerate which leads to additional complications as compared to the AdS case. Therefore, to keep things simple we are going to study states which are built acting on vacuum $\ket{0}$ with hard operators. We shall discuss the implications of our protocol for more general states towards the end of the paper.

The bulk state in general will be denoted by $\ket{\psi}$. All measurements are performed by observers who are located near $\scrip_-$. The observers are given two kinds of abilities:
\begin{enumerate}
 \item They can modify the state by acting on it with a unitary operator which has support on a small cut near $\mathcal{I}_{-}^{+}$.
 \item They are allowed to measure the energy of the state or a state modified by the action of a unitary\footnote{We assume that this measurement process does not induce a backreaction on the state.}.
\end{enumerate}
 We will prove below that having these two abilities are enough for the observers to determine the state completely. This will help establish a physical protocol via which one can, in principle, design experiments which demonstrate the principle of holography of information.

We pause to state an important point about causality. It might seem that causality is violated since the observers have access to all information by staying on a cut near $\scrip_-$. This is however not true since the observers are measuring the energy of the state which requires that they are spread across the full Celestial sphere (see eq.\eqref{energy}). Thus each observer only detects a part of the metric perturbation and they all have to meet at a point to sum their results in order to gain information about the energy of the state. Another way of thinking about this is to imagine that the observers have some kind of detectors which are {\it on} for a small time interval and switch {\it off} after they detect the gravitational radiation. Therefore, although the information is formally contained on a cut near $\scrip_-$, it is necessary for the observers to move out of that cut in order to physically reconstruct the state by assimilating the data on the detectors, which in general takes infinite time. This is consistent with the fact that the light crossing time in flat spacetime (time taken by light to reach null infinity) is infinite and hence there is no violation of causality. 

For convenience it is useful to work with a state that does not have an overlap with the vacuum $\ket{0}$. As has been shown in appendix B of \cite{Chowdhury:2020hse}, given the boundary values of any state $\ket{\psi}$, it is possible to find a unitary operator $V$ in the boundary algebra such that $\braket{0|V|\psi} = 0$. This means that instead of working with the state $\ket{\psi},$ we can always work with $V\ket{\psi}$ and establish the protocol to recover information for $V\ket{\psi}$. We shall not repeat the proof of this statement here and just state the physical intuition. By appropriately smearing the fields and their conjugate momenta it is possible to find a single mode near $\scrip_-$ which effectively behaves as a harmonic oscillator. The job of the observers then reduces to finding a single unitary matrix $V$, in this harmonic oscillator basis, which upon proper tuning ensures $\braket{0|V|\psi} = 0$. The method to construct such a matrix is explained in \cite{Chowdhury:2020hse}. Such a construction uses the entanglement of fields in the vacuum state. 
Henceforth, we shall assume that such a process has already been performed on a given state and $\ket{\psi}$ will denote states which do not have an overlap with the vacuum.

\subsection{Basis used for construction}\label{Basis for construction}
One crucial difference between the construction in AdS \cite{Chowdhury:2020hse} and flat spacetime is the absence of discrete energy eigenstates in the latter. This means that the energy eigenstates are a natural choice of basis for the expansion of the state in AdS but not in flat spacetime, as the energy is continuous. We therefore construct another basis called the {\it normal ordered basis} which allows us to reconstruct the state using a physical protocol\footnote{ The normal ordered basis can also be used in the AdS construction but we find that it is much more convenient to use the energy eigenstate basis in that case. It is important to note that this basis is formed out of continuous functions and there are certain subtle limitations in using this. These limitations are discussed in section \ref{discussion}.}. It will be shown how this method allows us to follow similar steps for reconstruction as those in the AdS case. For simplicity, we first explain the protocol by working with a state which is built out of operators of a single flavour. We later extend this to states built with multiple flavours.

Any state $\ket{\psi}$ constructed out of a single flavoured field $\phi$ on top of the vacuum $\ket{0}$ can be expanded in the normal ordered basis as
\begin{eqn}\label{psidefn1}
\ket{\psi} &= \int  \sum_{n = 1}^{\infty} \prod_{j = 1}^n : \phi(u_j, \Omega_j): g_n(\vec u, \vec \Omega) d \vec u d\vec \Omega \ket{0}
\end{eqn}
where $g_n(\vec u, \vec \Omega) \equiv g_n(u_1, \Omega_1; u_2, \Omega_2; \cdots; u_n, \Omega_n)$ are certain smooth smearing functions. Here $::$ denotes {\it normal ordering} and it is defined by pushing all the creation operators in the expansion to the left and the annihilation operators to the right\footnote{For example: $:\phi(u_1, \Omega_1) \cdots \phi(u_n, \Omega_n): \ket{0} \sim \int d\omega_1 \cdots d\omega_n e^{i \omega_1 u_1 + \cdots i \omega_n u_n}a^\dagg(\omega_1 \Omega_1) \cdots a^\dagg(\omega_n \Omega_n) \ket{0}$. The expansion of the field $\phi(u,\Omega)$ at $\scrip$ is derived in appendix \ref{app:saddle}.}. From the definition of the state $\ket\psi$ in \eqref{psidefn1} we see that $\braket{0|\psi} = 0$. We show in the following subsections how this choice of basis allows us to compute the functions $g_n$ in a {\it sieve procedure}, which means that $g_n$ can be evaluated only after obtaining $g_{n-1}$.

The task of the observers is to determine the function $g_n(\vec u, \vec \Omega)$ by performing certain kinds of measurements around $\scrip_-$. For example, the observers are allowed to measure conserved quantities like the Energy of the state. The observers are also allowed to manipulate the state by acting on it with a unitary operator located near a small cut at  $\scrip_-$. An expression for the energy in the Bondi gauge in $3+1$ dimensional flat spacetime is given as\footnote{A gauge invariant expression can be derived by following the procedure illustrated in \cite{Chowdhury:2021nxw}.} (this is equal to $\mathcal Q_{l =0, m =0}$ as defined in \eqref{supert})
\be\label{energy}
E = \frac{1}{16\pi G_N} \int_{\scrip_-} \sqrt{\gamma}d^2 \Omega \ m_B(u = -\infty, \Omega)\,.
\ee
The energy will be measured in a quantum sense with the energy of the vacuum state renormalized to $0$. We shall also assume that the Born-rule is valid for such measurements. This means that the answer to ``what is the frequency with which we obtain 0 upon measuring the energy of the state $\ket\psi$?'' is given as\footnote{In all our measurements, we will only be concerned with the frequency with which the energy is zero. One can also consider projectors onto an energy band close to zero, but as discussed in section \ref{discussion}, such modifications do not alter the result. This means that as long as the energy of the state is within a range $[0, \delta]$, it will be assumed to be the vacuum state. For non-zero energies away from $\delta$, such a measurement does not yield a useful result in flat spacetime as energy is continuous. }
\be
\braket{\psi|P_0|\psi}
\ee
where $P_0$ is the projector onto the vacuum state $\ket{0}$ as defined in \eqref{zeroprojector}. For states of the form shown in eq.\eqref{psidefn1} we clearly have $\braket{\psi|P_0|\psi} = 0$. Henceforth, when we write that we measure the energy of the state, we always mean a measurement of the kind above. Notice that in order to measure the energy of the state, the observers need to be spread across the entire Celestial sphere as each of them only measures a part of the metric fluctuation $m_B$ (which is suppressed by $G_N$, see eq.\eqref{ratemb}). This ensures that there is no violation of causality (see section \ref{discussion} for a discussion). 

The observers are also allowed to modify the state by acting on it with some unitary $U$, and then measure the energy of the modified state $U\ket{\psi}$. The unitaries that we will be using are of the form  
\be\label{unitary-1}
U_n = \exp\Big[i  \int_{-\infty}^{-1/\e} d\vec u' d\vec \Omega' \prod_{j = 1}^n f_n(\vec u', \vec \Omega' )O(u'_j, \Omega'_j) \Big]\,.
\ee
Here we denote the operators used by the observers with $O(u', \Omega')$ (although they are still the same field $\phi$) with $f_n(u', \Omega')$'s being smearing functions that localize these operators near $\scrip_-$, i.e, $u' \in (-\infty, -\frac{1}{\e}]$ . For these expectation values to be simple analytic functions it is useful to choose $O(u', \Omega') \sim u' \pi(u', \Omega')$, i.e, proportional to the conjugate momenta of the scalar fields. The factor of $u'$ is added for the sake of maintaining dimensions but can always be absorbed by an appropriate choice of $f$. 

We now show that a measurement of the form $\braket{\psi|U_n P_0 U_n|\psi}$ will allow us to fix the functions $g_n$ up to a phase factor. For this we first expand the unitaries up to the first order in $f$, i.e,
\be
U_n =  1 + i \int d\vec u d\vec \Omega \prod_{j = 1}^n f_n(\vec u', \Omega' )O(u'_j, \Omega'_j) + \mathcal{O}(f^2)\,. 
\ee
Henceforth, unless necessary, we shall suppress the $\mathcal{O}(f^2)$ terms in the expressions below. 

Let us consider the measurement where we compute the energy of the state $U_1\ket{\psi}$ and compute the frequency with which we get zero.  By the Born-rule, this is equivalent to computing $\braket{\psi|U_1^\dagg P_0 U_1|\psi}$. This correlator is equal to (we refer the reader to appendix \ref{app:softstructure} for the details of this computation)
\begin{eqn}
 \braket{\psi|U_1^\dagg P_0 U_1|\psi} &= \Big| \int du du' d\vec \Omega d\vec \Omega' \ f_1(u', \Omega') g_1(u, \Omega) \braket{0|\phi(u, \Omega) O(u_j', \Omega_j')|0} \Big|^2\,.
\end{eqn}
The correlation function above allows us to determine the function $g_1(u, \Omega)$ up to a phase factor, which we denote by $e^{i \q_1}$
\bes
 \int du du' d\vec \Omega d\vec \Omega' \ f_1(u', \Omega') g_1(u, \Omega) \braket{0|\phi(u, \Omega) O(u_j', \Omega_j')|0} = \sqrt{\braket{\psi|U_1^\dagg P_0 U_1|\psi}} e^{i \q_1}\,.
\ees
In appendix \ref{app:recovery} we explain how the function $g_1$ can be reconstructed from such an integral equation. Since the overall phase of the state $\ket\psi$ is not a physically measurable quantity, we can choose it such that $e^{i \q_1} = 1$. This integral equation completely fixes the function $g_1$ for us\footnote{It is useful to contrast the operator $U_1$ with the operator $X_r$ defined (on page 10) in \cite{Chowdhury:2020hse}.}. We refer the reader to appendix \ref{app:recovery} for further details. 

In the following subsection \ref{sec:correlatorrecovery} we explain how we obtain the functions $g_n$ when $n\neq 1$ using appropriate correlation functions. Subsequently in subsections \ref{sec:singleflavourprotocol}-\ref{sinthetagen} we develop a physical protocol that a set of observers near $\scrip_-$ can use to recover the states.

\subsection{Information recovery using correlation functions}\label{sec:correlatorrecovery}
We demonstrate a simple use of the normal ordered basis \eqref{psidefn1} by allowing ourselves to measure arbitrary expectation values. We show how one can easily obtain all $g_{n}$'s having determined $g_1$, by measuring specific correlation functions. We note that this procedure has to be performed in a sieve-like manner, i.e, we can determine the value of $g_n$ once we know the value of $g_{n-1}$. Let us consider the following correlation function $\braket{\psi|U_1^\dagg P_0 U_n|\psi}$ at $O(f^2)$
\begin{eqn}\label{correlationfuncsingle}
 \braket{\psi|U_1^\dagg P_0 U_n|\psi} &=\int d\vec u d \vec u' d\vec \O d\vec \O' du du' d\Omega d\Omega' \sum_{j = 1}^n f_1(u', \Omega') f_n(\vec u', \vec \Omega') g_1^*(u, \Omega) g_j(\vec u, \vec \Omega) \\ 
 &\qquad\times \braket{0|\phi(u, \Omega)O(u', \Omega')|0} \braket{0|O_1 \cdots O_n :\phi_1 \cdots \phi_j:|0}
\end{eqn}
where we use the shorthand notation $O_n \equiv O(u_1', \Omega_1'; \cdots; u_n', \Omega_n')$, $\phi_n \equiv \phi(u_1, \Omega_1; \cdots; u_n, \Omega_n)$ and $g_n(\vec u, \vec \Omega)\equiv g_n(u_1, \Omega_1; \cdots; u_n, \Omega_n)$. Therefore upon measuring $\braket{\psi|U_1^\dagg P_0 U_n|\psi}$ for all $n$, starting with $n = 2$, we easily obtain the value for all $g_{n> 1}(\vec u, \vec\Omega)$.

Such a measurement is not physically viable as the final answer is not real in general. However, it demonstrates a simple use of the normal ordered basis in order to extract information about the state in a sieve-like manner. It has been argued in \cite{Laddha:2020kvp} that the measurement of all possible correlation functions allows a complete reconstruction of the state and hence it was expected that such a procedure should exist.

\subsection{Physical Protocol}\label{sec:singleflavourprotocol} 
In the following section we explain how we can obtain the functions $g_n$ by performing measurements that are physically viable.

As explained in the subsection \ref{Basis for construction}, we can obtain the function $g_1$ by measuring $\braket{\psi|U_1^\dagg P_0 U_1|\psi}$ and  exploiting the freedom to choose the overall phase of the state $\ket{\psi}$. Since we can only fix the overall phase of the state once, this procedure will still leave a phase ambiguity $e^{i \q_n}$ for all other $g_{n > 1}$. To see this ambiguity explicitly, we first modify the state by acting on it with a unitary $U_n$, then measure its energy and see the frequency with which we get $0$. Using the Born-rule this is equivalent to measuring $\braket{\psi|U_n^\dagg P_0 U_n|\psi}$. Upon expanding this to $\mathcal{O}(f^2)$ we get, 
\be
\braket{\psi|U_n^\dagg P_0 U_n|\psi} = \Big| \braket{0|\int d \vec u' d \vec \Omega'  f_n(\vec u', \vec \Omega') O_1 \cdots O_n |\psi}  \Big|^2 \,.
\ee
By inverting this relation we obtain the correlation function which characterizes the phase ambiguity $e^{i \q_n}$
\be
\braket{0|\int d \vec u' d \vec \Omega'  f_n(\vec u', \vec \Omega') O_1 \cdots O_n |\psi} = \sqrt{\braket{\psi|U_n^\dagg  P_0  U_n|\psi}} e^{i \q_n} \,.
\label{thetadefn}
\ee
The advantage of expanding $\ket\psi$ in the normal ordered basis \eqref{psidefn1} becomes obvious in this step; using such an expansion ensures that the only $g_i$'s contributing to the correlator on the LHS are for $i \leq n$. Hence we have to evaluate these correlators in a sieve-like procedure since only after one determines the value of $g_n$, one can determine $g_{n+1}$.

We will now demonstrate how the phase ambiguities $e^{i \q_n}$ can be fixed by making a two-step measurement. This requires the action of two unitary operators on the state and then measuring its energy. This results in correlation functions of the form $\braket{\psi|U_j^\dagg U_i^\dagg P_0 U_i U_j|\psi}$. In the following subsections we explain how this fixes the value of $\q_n$ completely, by first determining $\cos\q_n$ and then the value of $\sin\q_n$. 

\subsection{Determining $\cos\q_n$}
As shown above, the phase of $g_1$ is completely fixed by making a choice for the overall phase of the state $\ket{\psi}$. This will allow us to compute the value of $\cos\q_n$ by performing a two-step measurement of the form $\braket{\psi|U_n^\dagg U_1^\dagg P_0 U_1 U_n|\psi}$ at $\mathcal{O}(f^2)$. A simple calculation shows

\begin{eqn}\label{costheta-n}
 \cos\q_n = \frac{ \braket{\psi|U_1^\dagg U_n^\dagg P_0 U_n U_1|\psi} - \braket{\psi|U_1^\dagg P_0 U_1|\psi}- \braket{\psi|U_n^\dagg P_0 U_n|\psi}}{2\sqrt{\braket{g|U_1^\dagg P_0 U_1|g}} \sqrt{\braket{g|U_n^\dagg P_0 U_n|g}}}\,. 
\end{eqn}
However this does not completely fix $\q_n$ since determination of $\cos\q_n$ leaves us with an ambiguity for the sign of $\sin\q_n$. In the following subsection we explain how we can fix the value of $\sin\q_n$ for all $n$.

\subsection{Determining $\sin\q_n$}\label{sinthetagen}
In order to fix $\sin\q_n$ we just need one $g_n$ whose phase $e^{i\q_n}$ is not purely real. The fact that the phase of the function $g_1$ was chosen to be purely real allowed us to measure the value of $\cos\q_n$. However in general we do not expect the phases of all $g_n$ to be purely real. This can be checked by evaluating the value of $\cos\q_n$ using eq.\eqref{costheta-n} and as long as for some $n = n_0$, $\cos\q_{n_0} \neq \pm 1$, we can perform an analogues two-step measurement involving $U_{n_0}$ to determine the sign of $\sin\q_n$. Such an $n_0$ can be easily obtained by trial and error. Then, we can perform a measurement of the form $\braket{\psi|U_{n_0}^\dagg U_n^\dagg P_0 U_n U_{n_0}|\psi}$ at $\mathcal{O}(f^2)$
\begin{eqn}\label{allsigns}
 \braket{\psi|U_n^{\dagg} U_{n_0}^{\dagg} P_0 U_{n_0} U_n|\psi} &= \braket{\psi|U_n^\dagg P_0 U_n^\dagg|\psi} + \braket{\psi|U_{n_0}^\dagg P_0 U_{n_0}^\dagg|\psi} \\
 &\quad + 2\sqrt{\braket{\psi|U_n^\dagg P_0 U_n|\psi}} \sqrt{\braket{\psi|U_{n_0}^\dagg P_0 U_{n_0}|\psi}}  \big( \cos\q_{n} \cos\q_{n_0} + \sin\q_{n}\sin\q_{n_0}\big)\,.
\end{eqn}
In this measurement we end up with a correlated ambiguity in the phases of $g_n$ and $g_{n_0}$, which implies that upon knowing the value of $\sin\q_{n_0} \neq 0$ for any given $n_0$, we can easily determine the value of all other $\sin\q_n$. We now explain how we can fix the value of $\sin\q_{n_0}$.

\subsubsection{Determining the sign of $\sin\q_{n_0}$}\label{sinn0}
As shown in the previous section, by performing certain simple measurements up to $\mathcal{O}(f^2)$, we decode a lot of information about the state $\ket{\psi}$. However, we are left with one final sign ambiguity concerning $\sin\q_{n_0}$. Since this is only one sign ambiguity, we just need one measurement which can distinguish between $e^{i \q_{n_0}}$ and $e^{- i \q_{n_0}}$.

We shall work with the special case $n_0 = 2$ to demonstrate the procedure, but the method can be performed for a generic $n_0$ as well. 
Consider a measurement of the kind $\braket{\psi|U_1^\dagg P_0 U_1|\psi}$, upon expanding it up to $\mathcal{O}(f^3)$ we get
\begin{eqn}\label{sinn0-1}
 \braket{\psi|U_1^\dagg P_0 U_1|\psi} &= \int du_1' du_2'  d\Omega_1' d\Omega_2'  f_1(u_1', \Omega'_1) f_1(u_2', \Omega'_2)  \braket{\psi| O(u_1', \Omega_1') P_0 O(u_2', \Omega_2')|\psi} \\
 &+ i \int d\vec u'd\vec \Omega' f_1(u_1', \Omega_1') f_1(u_2', \Omega_2')f_1(u_3', \Omega_3') \Big[ \braket{\psi|O_1 P_0 O_2 O_3|\psi} - \braket{\psi|O_2 O_3 P_0 O_1|\psi}  \Big]
\end{eqn}
where $d\vec u' d\vec \Omega' = du_1' du_2' du_3' d\Omega_1' d\Omega_2' d\Omega_3'$ and $O_i = O(u_i', \Omega_i')$. As explained in the sections before, the terms at $\mathcal{O}(f^2)$ enable us to completely fix the function $g_1(u, \Omega)$. Let us now focus on the term at $\mathcal{O}(f^3)$ and expand $\ket\psi$ using \eqref{psidefn1},
\begin{eqn}
&i \int d\vec u'd\vec \Omega' f_1(u_1', \Omega_1') f_1(u_2', \Omega_2')f_1(u_3', \Omega_3') \Big[ \braket{\psi|O_1 P_0 O_2 O_3|\psi} - \braket{\psi|O_2 O_3 P_0 O_1|\psi}  \Big]\\
&= i\int d\vec u'd\vec \Omega' d u d\Omega d\vec u d\vec \Omega \ f_1(u_1', \Omega_1') f_1(u_2', \Omega_2')f_1(u_3', \Omega_3')  g_1(u, \Omega)\\
&\quad \times \Big[ g_2(\vec u, \vec \Omega) \braket{0|\phi(u, \Omega) O_1|0\rangle \langle 0|O_2 O_3 :\phi_1 \phi_2: |0} - g_2^*(\vec u, \vec \Omega) \braket{0|:\phi_1 \phi_2:O_2 O_3 |0\rangle \langle 0| O_1 \phi(u, \Omega)|0}  \Big]
\end{eqn}
where we have used the notation $\phi_n = \phi(u_n, \Omega_n)$. Such a measurement does not allow us to reconstruct the full function $g_2$ since the only localizing functions appearing in this expression are dependent on $f_1(u', \Omega')$. However it is still sensitive to the sign of $\sin\q_2$ since (see appendix \ref{app:saddle} for details on evaluating such correlators)
\be
\braket{\psi|U_1^\dagg P_0 U_1|\psi} \mbox{ at $\mathcal{O}(f^3)$ }\sim (g_2 - g_2^*)  \sim \sin\q_2\, .
\ee
Upon determining the sign of $\sin\q_2$ we can determine all the other signs using \eqref{allsigns}. Such a construction can be easily extended for any $n_0 \neq 2$ as well. This allows us to determine the value of $\sin\q_{n_0}$ and therefore fix the full wave function $\ket\psi$. 

Notice that for this proof we are assuming that we have access to both even and odd $g_n$'s, i.e, we need at least one of each $g_{odd}$ and $g_{even}$ $\neq 0$. This excludes the two special cases when the only non-zero $g_n$'s are either made out of all even or all odd $n$. These cases can be dealt with in a similar manner as above but with a slight difference and are presented in appendix \ref{app:oddeven}.

\subsection{Multiple flavors}\label{sec:multipleflavours}
We now explain how the results of the previous section can be extended to cases when we have multiple flavours. To keep things simple, we illustrate the algorithm when we have two flavours only, but it can be trivially generalized for multiple flavours. 

Consider the following state
\begin{eqn}
 \ket{\psi} &=  \int  d\vec u d\vec \Omega \Big[ - g_{00} + \sum_{i = 0}^{\infty} \sum_{j = 0}^{\infty} g_{ij} :\phi_1 \cdots \phi_i: \ :\tilde \phi_1 \cdots \tilde\phi_j: \Big]  \ket{0} 
\end{eqn}
where the term $-g_{00}$ ensures that $\braket{0|\psi} = 0$ and $\phi$, $\tilde \phi$ are two different flavours of scalar fields. We have omitted the dependence of $g_{ij}$ and $\phi, \tilde\phi$ on $(u, \Omega)$ to avoid a clutter of notation. In a manner similar to the previous sections, we shall determine the values of all $g_{ij}$ in a sieve-like procedure. The unitary operator used to modify the state is denoted as, 
\be
U_{ij} = \exp\Big[i \int d\vec u' d\vec \Omega' \ f_{ij}(\vec u', \vec \Omega') O_1 \cdots O_i \tilde O_1 \cdots \tilde O_j\Big]
\ee
where $O$ and $\tilde O$, describe the same fields as $\phi$ and $\tilde \phi$, but they are localized near $\scrip_-$ (as ensured by $f_{ij}$).

We first explain how the functions $g_{ij}$ are recovered using arbitrary correlation functions and then go on to demonstrate the same using a physical protocol. The steps performed are very similar to the ones in the case of a single flavour. 

\subsubsection{Arbitrary correlators}
Here we extend the procedure presented in section \ref{sec:correlatorrecovery} to determine the functions $g_n$ for multiple flavours. We first have to measure the value of $g_{10}$ by measuring $\braket{\psi|U_{10}^\dagg P_0 U_{10}|\psi}$ at $\mathcal{O}(f^2)$.  This allows us to measure $g_{10}$ up to a phase $e^{i \q_{10}}$. This phase $\q_{10}$ can be set to zero by exploiting the fact that the overall phase of the state $\ket \psi$ is not a physically measurable quantity. 

Upon determining $g_{10}$ we can easily determine all the other $g_{ij}$'s in a sieve-like manner by evaluating the following correlation function
\begin{eqn}
&\braket{\psi|U_{10}^\dagg P_0 U_{ij}|\psi}\\
&= \int dV f_{10} f_{ij} g_{10} \sum_{m = 0}^i \sum_{n = 0}^j g_{mn} \braket{0|\phi_1 O_1|0} \braket{0|O_1 \cdots O_i \tilde O_i \cdots \tilde O_j :\phi_1 \cdots \phi_m: :\tilde \phi_1 \cdots \tilde \phi_n:  |0}
\end{eqn}
with $dV = d\vec u d\vec u' d\Omega d\Omega'$ and we use the same shorthand notation for $\phi_i$, etc. as defined earlier. Clearly this is not a physical protocol since the correlation functions in general are not real. However it demonstrates the use of the normal ordered basis to decode generic states. In the following section we explain how one can determine the value of the smearing functions by using a physical protocol.

\subsubsection{Physical Protocol}
We now explain how the analysis in the previous section can be extended to a physical protocol where the only kind of measurements allowed are the ones which give real answers. For a single flavour this is explained in section \ref{sec:singleflavourprotocol}. Due to the conceptual similarity between the single and the multiple flavoured case it suffices to summarize the main steps without repeating the details. We remind the reader that this entire procedure is performed in a sieve-like manner, i.e, one should have measured the value of $g_{mn}$ before determining $g_{i>m, j>n}$.

\begin{enumerate}
 \item We first fix the value of $g_{10}$ by exploiting the fact that the overall phase of the wave function is meaningless. This is exactly similar to fixing the value of $g_1$ in the single flavour case. 
 
 \item The general phase ambiguity in the function $g_{ij}$ can be quantified by the measurement at $\mathcal{O}(f^2)$ of
 \be
 \braket{s|\int f_{ij} O_1 \cdots O_i \tilde O_1 \cdots \tilde O_j|\psi} =\sqrt{\braket{\psi|U_{ij}^\dagg P_0 U_{ij}|\psi}} e^{i \q_{ij}}\,.
 \ee
 The only phase fixed till now is $\q_{10} = 0$. 
 
 \item By measuring $\braket{\psi|U_{ij}^\dagg P_0 U_{ij}|\psi}$ we obtain the value of $g_{ij}$ up to a phase factor $e^{i \q_{ij}}$.

 \item  The value of $\cos\q_{ij}$ can be fixed by measuring $\braket{\psi|U_{ij}^\dagg U_{10}^\dagg P_0 U_{10} U_{ij}|\psi}$ at $\mathcal{O}(f^2)$. This uses the fact that we have fixed $\q_{10}= 0$. Fixing the value of $\cos\q_{ij}$ does not completely fix $\q_{ij}$ since it leaves us with a sign ambiguity for $\sin\q_{ij} \ \forall \ i,j$.

 \item In order to fix the value of $\sin\q_{ij}$ for all $i, j$, we first need to fix the value of $\sin\q_{ij}$ for some particular $i, j$. This can be done (for example) for $i = 2, j = 0$ by measuring $\braket{\psi|U_{10}^\dagg P_0 U_{10}|\psi}$ at $\mathcal{O}(f^3)$. Noting the analogy with the single flavoured case, the value of other $\sin\q_{ij}$ can be fixed by measuring $\braket{\psi|U_{20}^\dagg U_{ij}^\dagg P_0 U_{ij} U_{20}|\psi}$ $\forall \ i, j$. 
\end{enumerate}
The special cases where we have all even or all odd $g$'s are similar to those of the single flavoured case and are shown in appendix \ref{app:oddeven}.

\section{Conclusion and Discussion}\label{discussion}

\subsection*{Main Result}

We presented a physical protocol that observers localized on a cut near $\scrip_-$, in a theory of quantum gravity coupled to massless matter in $3+1$ dimensional asymptotically flat spacetime, can use to detect bulk massless excitations on a given vacuum state $\ket{0}$. The protocol comprises the manipulation of the state with unitary operators and the measurement of its overlap with the vacuum. These measurements are performed in a manner which is quantum mechanical in nature and utilizes the Born-rule. 
The protocol presented in this paper is an extension of a similar result for asymptotically AdS spacetimes \cite{Chowdhury:2020hse} . 
The crucial difference between the two is of a technical nature. In the AdS example we had used the basis of energy eigenstates to expand any given state and this step was crucial for developing the protocol. Unlike AdS, the energy spectrum in flat spacetime is continuous, thus the energy eigen basis (which can be thought of as a band of energy) is not a convenient basis to expand the states. To overcome this difficulty, we introduced a normal ordered basis \eqref{psidefn1} to expand the states. Then the observers that are localized at a small cut near $\scrip_-$ can extract information in a sieve-like procedure.

\subsection*{Limitation of the Normal Ordered Basis}
At this point, we would like to discuss the main limitation of using the normal ordered basis in our construction. The state $\ket\psi$ when expanded in the normal ordered basis \eqref{psidefn1} is defined using smearing functions $g_n(\vec u, \vec \Omega)$ which are continuous. As shown in appendix \ref{app:recovery}, these functions are reconstructed from their moments. In principle, one needs infinitely many moments to reconstruct a function exactly, however, in any physical experiment we only have access to a finite number of moments.  This introduces a natural cut-off in the accuracy with which the smearing functions can be reconstructed. For example, the protocol described in the main text fails when the smearing functions are highly oscillatory, i.e, non-smooth. This is because, it will not be possible for the observers to distinguish between two smearing functions which differ beyond a certain moment.

\subsection*{Effect of Noise}
In all the measurements we assume that the energy of the vacuum state is a known quantity and is renormalized to zero. Therefore in most measurements that we perform, we end up with the projector onto the vacuum state $P_0$. However in general one can also consider  projectors onto a small but finite energy $P_\delta$ and ask if the results change as we take $\delta \to 0$. It is simple to see that upon normalizing the projector correctly one does not end up getting different results from the ones obtained by using $P_0$. Thus, as long as the energy of a state is in between $[0, \delta]$, the state is assumed to be the vacuum state. Such projectors have also been considered in \cite{Chakraborty:2021rvy} in relation to the monogamy paradox in flat spacetime.

\subsection*{Construction vs Reconstruction}
By using the Reeh-Schlieder theorem \cite{Laddha:2020kvp, Haag:1992hx, Witten:2018zxz} it is possible to show that any state can be created by acting with hermitian operators, localized near $\scrip_-$, on the vacuum $\ket{0}$\footnote{More generally this construction works by acting with an operator on a state which is cyclic and separating.}. This is also possible in AdS \cite{Chowdhury:2020hse, Banerjee:2016mhh} where any state can be created by acting with operators, localized on a thin time band near the boundary, on the vacuum. Such a construction only requires the positivity of energy as measured from asymptotic infinity and is valid for non-gravitational QFTs as well.

However note that the process of detecting a state by being able to perform measurements only using operators localized at a small cut near the boundary is a non-trivial process as compared to creating the state. Reeh-Schlieder like constructions do not guarantee that the state can be reconstructed by performing measurements near the boundary of spacetime (both in flat spacetime and in AdS). One simple way to see this is by considering the case of a local QFT in flat spacetime. Even in this case, it is possible to construct any state of the QFT by acting with operators (near $\scrip_-$) on the vacuum. However, as argued in the main text, it is clearly not possible to reconstruct this state exactly since there exists local operators in the bulk which commute with observables at the boundary. Therefore, although it is typically possible to construct any state in flat spacetime by acting with operators near $\scrip_-$, reconstructing the state is only possible in a quantum theory of gravity.

\subsection*{Locality and Causality}
One of the main reasons we are able to reconstruct information about a general bulk state from a cut near $\scrip_-$, is because the theory of quantum gravity is not strictly local. This non-locality arises because of the primary and secondary constraints in gravity. It was shown in \cite{Chowdhury:2021nxw} that for wave functionals satisfying the Wheeler-DeWitt constraints in perturbation theory about global AdS, one can obtain a notion of holography. Similar constraints also exist in a gauge theory, like QED, but with a different structure for the constraints, which allows the formation of {\it split states} in QED but not in quantum gravity \cite{Raju:2021lwh}. The presence of such states prevents the reconstruction of information in a gauge theory. It will be interesting to see how such results generalize to flat spacetime.

As explained in the main text, such a protocol for reconstruction does not lead to violation of causality. This is because the observers making measurements are spread across the Celestial sphere near $\scrip_-$ and have to meet at some point to sum up their results. Clearly this will take more than light crossing time (which itself is infinite in flat spacetime). Thus the information being  formally contained in $\scrip_-$ does not lead to any violation of causality as the observers always need more than light crossing time to reconstruct the information they measure. 

\subsection*{Detection of Memory Effect \& States Built on Multiple Vacua}
The protocol established in this paper gives a procedure on how to determine hard excitations on a given vacuum state. However it is possible to consider states which are built on multiple vacuum states, for example,
\bes
\ket \Psi \equiv \sum_{m = 1}^M\sum_{n = 1}^{\infty} \prod_{j = 1}^n \int  : \phi(u_j, \Omega_j): g_n^{(m)}(\vec u, \vec \Omega) d \vec u d\vec \Omega \ket{0_{(m)}}~.
\ees
The main difficulty in decoding states of this kind is that the observers near $\scrip_-$ do not apriori know which vacuum states $\ket{0_{(m)}}$ are used to build $\ket{\Psi}$. If the observers are given priors about the particular $\ket{0_{(m)}}$'s that appear, they can use certain special unitaries constructed out of the Super-rotation charge \cite{Compere:2016jwb} in order to bring the state $\ket\Psi$ to the form of \eqref{psidefn1}. Without that information, it is not possible to detect the state using the protocol illustrated in the main text. In order to detect these states in the absence of priors, one would have to make use of the transition operators defined in \cite{Laddha:2020kvp}. However, it is not clear how one can measure such transition operators  in a physical process.

There is a similar restriction for detecting the Memory effect \cite{Strominger:2014pwa, Satishchandran:2019pyc} as it concerns the transition from a given configuration of vacuum to another one. Such transitions typically occur in scattering processes from $\scrim$ to $\scrip$ but it is not a priori clear how one can study them by performing measurement only at $\scrip$ (or $\scrim$), we hope to address this in a future work.

\subsection*{Generalizations to Higher dimensions \& Massive fields}
Due to the similarities between supertranslations in four dimensions and other higher even dimensions \cite{Aggarwal:2018ilg, Pate:2017fgt, Chowdhury:2022nus} our results should easily generalize to higher even dimensions. However, generalizing this result to odd dimensions has to be explored in greater detail. Additionally, it would be interesting to extend such a protocol in the presence of massive fields and establish the principle of holography of information in that context.

\section*{Acknowledgements}
We are grateful to Suvrat Raju for suggesting this problem and his guidance throughout the project. We sincerely thank Panos Betzios, Nava Gaddam and Suvrat Raju for their comments on the draft. We also thank Anupam A. H., Panos Betzios, Tuneer Chakraborty, Joydeep Chakravarty, Nava Gaddam, Victor Godet, Alok Laddha, Ruchira Mishra, Priyardarshi Paul, Siddharth Prabhu, Ana-Maria Raclariu, Pushkal Shrivastava and Spenta Wadia for various discussions. C.C. acknowledges support from the Department of Atomic Energy, Government of India, under project no. RTI4001. O.P. acknowledges support from the Simons foundation. Research at Perimeter Institute is supported in part by the Government of Canada through the Department of Innovation, Science and Economic Development and by the Province of Ontario through the Ministry of Colleges and Universities.

\begin{appendix}

\section{Soft Structure of Vacuua}\label{app:softstructure}
The correlation functions that we measure are of the form $\braket{\psi|U_i^\dagg P_0 U_j|\psi}$ where $P_0$ is defined in eq.\eqref{zeroprojector}, the states $\ket{\psi}$ are generically of the form eq.\eqref{psidefn1} and the unitaries $U$ are of the form eq.\eqref{unitary-1}. The following discussion only refers to states built out of a single flavoured field and the extension to multiple flavours is similar. Since the states and the unitaries are constructed out of hard operators only (i.e, do not include operators like the zero mode of the shear), we show that the soft structure of the vacuum and the projector become unimportant\footnote{We thank Tuneer Chakraborty and Priyadarshi Paul  for several useful discussions on this and clarifying many doubts.}. From the decomposition of the vacuum and the projector as given in \eqref{vacuum} and \eqref{zeroprojector}, a generic correlation function of interest takes the form 
\be\label{softvac-1}
\braket{\psi|U_i^\dagg P_0 U_j|\psi} = \braket{0|Z^\dagg U_i^\dagg P_0 U_j Z|0} = \int Ds Ds' Ds'' g_{\{s\}} g_{\{s''\}}\braket{\{s\}|Z^\dagg U_i^\dagg |\{s'\}} \braket{\{s'\}|U_j Z|\{s''\}}
\ee
with $Z$ representing the operators on the RHS in the definition of the state \eqref{psidefn1} and $Ds \equiv \prod_{lm}ds_{lm}$. To avoid cluttering of notation we suppress the $l,m$ indices but it is trivial to reinstate them. We now use the fact that \cite{Ashtekar:2018lor}
\be
\braket{\{s\}| O|\{s'\}} = \braket{0| O|0} \delta(\{s\} - \{s'\})
\ee
for any hard operator $O$. This can be intuitively understood by noting that the operator $O$, a hard operator, does not induce a vacuum transition as it does not contain the zero mode of the shear. The above result can be also proven explicitly by using the condition $\braket{0|0} =1$. Using this result we can simplify \eqref{softvac-1} to get
\begin{align}
\braket{\psi|U_i^\dagg P_0 U_j|\psi} &= \int Ds Ds' Ds'' g^*_{\{s\}} g_{\{s''\}}  \braket{0|Z^\dagg U_i^\dagg|0} \braket{0|U_j Z|0} \delta(\{s\} - \{s'\})  \delta(\{s'\} - \{s''\}) \nonumber\\ 
&= \braket{0|Z^\dagg U_i^\dagg|0} \braket{0|U_j Z|0} \int |g_{\{s\}}|^2 \,Ds = \braket{\psi|U_i^\dagg|0} \braket{0|U_j|\psi}
\end{align}
This is an expected identity when there is no vacuum degeneracy. For this identity to hold in the presence of a degeneracy it is necessary that we are not considering expectation values of soft operators (like the zero mode of the shear).

 \section{Saddle point approximation for scalar field}\label{app:saddle}
In this appendix we work out the 2-point Wightman function for a free massless scalar field at null infinity in $(d+2)$-dimensional Minkowski spacetime. The free field correlation functions suffice at leading order in large$-r$ as the interaction terms do not survive at this order. A similar derivation can be performed for gauge and gravitational fields as well \cite{Strominger:2017zoo}. 

The mode expansion for the scalar field in the bulk is given as
\be
\phi(t, r, \Omega) = \int \frac{d^{d+1} q}{(2\pi)^{d+1}} \frac{1}{2\omega_q} \big[ a_{q} e^{i q \cdot x}  + a_{q}^{\dagg} e^{-i q \cdot x}  \big].
\ee
 The mode expansion at null infinity can be obtained by taking  the large$-r$ limit of this expression. It it convenient to decompose $e^{iq\cdot x} = e^{- i \omega u} e^{-i \omega r (1 - \cos\q)}$,
where we use $|\vec q| = q^0 = \omega$ and $\q$ denotes the angle between $\vec q$ and $\vec r$. In the limit $r \to \infty$ the only saddle that contributes is $\q = 0$ (the saddle at $2\pi$ is prevented by the Riemann-Lebesgue lemma). This allows us to approximate $1 - \cos \q \approx \frac{\q^2}{2}$ and therefore $\lim_{r \to \infty}e^{i q \cdot x} = e^{- i \omega u} e^{-i \omega r \q^2/2}$. 

Using this we perform the $\int d^{d+1} q$ integral (where the integral over the remaining $(d-1)$ variables is denoted by $S_{d-1}$.),
\be
\int \frac{d^{d+1}q}{2\omega_q} a_q e^{i q \cdot x} = -\frac{S_{d-1}i}{2r} \int d\omega \omega^{d-2}  a(\omega\hat \Omega) e^{-i \omega u} 
\ee
Thus the expansion of the scalar field at $\scrip$ using the saddle point approximation becomes
\be
\phi(u, \Omega) = - \frac{i S_{d-1}}{2(2\pi)^{d+1}} \intsinf d\omega \omega^{d-2}\big[ a(\omega \hat \Omega) e^{- i \omega u} - a^\dagg(\omega \hat \Omega) e^{i \omega u}  \big]
\ee
For computing the 2-point Wightman function of the fields $\braket{s|\phi(u, \Omega) \phi(u', \Omega')|s}$, it is necessary to compute the expectation values of the ladder operators $a, a^\dagg$.  This can be obtained by computing the commutator of the ladder operators
 \be
 [a(\omega \hat \Omega), a^\dagg(\omega' \hat \Omega')] = \frac{2}{\omega^{d-1}}(2\pi)^d \delta(\omega - \omega') \delta^d(\Omega,\Omega').
 \ee
 Since $a \ket{0} = 0$ a simple computation then gives us the 2-point Wightman function of the scalar field
\begin{eqn}\label{propagator}
\braket{0|\phi(u, \Omega) \phi(u', \Omega')|0} &=  \frac{S_{d-1}^2 i^{2-d} \Gamma(d-2) }{(2\pi)^{2d+2}} \times \frac{\delta^d(\Omega, \Omega')}{(u - u' - i \e)^{d-2}} \,,
\end{eqn}
where we have introduced a factor of $i\e$, with $\e > 0$, for the integral to converge. In $d=2$ this gives a Logarithmic dependence and it is therefore more convenient to evaluate $\braket{0|\pi(u, \Omega)\phi(u', \Omega')|0} = \braket{0|\p_u\phi(u, \Omega)\phi(u', \Omega')|0} $,
\begin{eqn}\label{phipi-1}
 \braket{0|\pi(u, \Omega)\phi(u', \Omega')|0}_{d = 2} = - i \frac{ \delta^2(\Omega, \Omega')}{4\pi} \intsinf d\omega e^{-i \omega(u - u' - i \e) } = -\frac{1}{4\pi} \frac{\delta^2(\Omega, \Omega')}{u - u' - i\e} 
\end{eqn}
where we use $S_1 = 2 \pi$.

 \section{Explicit reconstruction of $g_n$'s}\label{app:recovery}
In this appendix we explain how one can explicitly reconstruct the functions $g_n$'s. The kind of correlation functions that we encounter lead to equations of the following kind for $g_n(\vec u, \vec \Omega)$
\be
\int d\vec u d\vec\Omega d\vec u' d\vec \Omega' \ f_n(\vec u', \vec \Omega') g_n(\vec u, \vec \Omega) \frac{\delta(\Omega_1, \Omega'_1)}{u_1 - u'_1 - i \e} \cdots \frac{\delta(\Omega_n, \Omega'_n)}{u_n - u'_n - i \e} = C_n.
\ee
Here  $f_n(\vec u, \vec\Omega')$ is a smearing function localized near $\scrip_-$ which is up to the observers to tune and $C_n$ is the result of the measurement they perform. The structures $\frac{\delta(\Omega, \Omega')}{u - u' -i \e}$ arise from the 2-point functions computed in appendix \ref{app:saddle}. The integration of the delta functions gives
\be
\int d\vec u d\vec\Omega d\vec u' \ f_n(\vec u', \vec \Omega) g_n(\vec u, \vec \Omega) \frac{1}{u_1 - u'_1 - i \e} \cdots \frac{1}{u_n - u'_n - i \e} = C_n\,.
\ee
From this equation we see that the spherical dependence of $g_n$'s is fixed as the functions $f_n$'s can be tuned arbitrarily by the observers. In order to fix the $u$ dependence, let us consider the integrand of $\vec u'$ on the LHS
\be
\mathcal I \equiv \int d\vec u g_n(\vec u, \vec \Omega) \frac{1}{u_1 - u'_1 - i \e} \cdots \frac{1}{u_n - u'_n - i \e} 
\ee
and expand the integrands as $u'_i \to -\infty$ (since these arise from fields that are localized near $\scrip_-$) which gives

\be
\mathcal I = \sum_{n_1, \cdots, n_n = 0}^\infty \frac{(-1)^{n\mbox{\scriptsize{mod}} 2}}{u_1'^{n_1+ 1} \cdots u_n'^{n_n+1}} \int  g_n(\vec u, \vec \Omega) u_1^{n_1} \cdots u_n^{n_n} d\vec u\,.
\ee
Plugging this integrand back into $C_n$ we obtain 
\be
\int d\vec u' d\Omega f(\vec u', \vec \Omega) \sum_{n_1, \cdots, n_n = 0}^\infty \frac{(-1)^{n\mbox{\scriptsize{mod}} 2}}{u_1'^{n_1+ 1} \cdots u_n'^{n_n+1}} \int  g_n(\vec u, \vec \Omega) u_1^{n_1} \cdots u_n^{n_n} d\vec u = C_n \,.
\ee
Since the observers are allowed to tune the function $f_n(\vec u', \vec \Omega)$ arbitrarily, they can obtain all possible moments of the distribution $g_n(\vec u, \vec \Omega)$ and therefore reconstruct the function $g_n(\vec u, \vec \Omega)$ completely. We have included a mathematica notebook with this submission that demonstrates how this works in a numerical example for a 1-dimensional function.

 \section{Special cases of states}\label{app:oddeven}
In this appendix we explain how we can determine the state when we have only even or only odd excitations. The states we consider are of the form 
\begin{subequations}
\be\label{even}
 \ket{\psi_{even}} = \sum_{n = 1}^{\infty} \prod_{j = 1}^{2n} \int  : \phi(u_{j}, \Omega_{j}): g_{2n}(\vec u, \vec \Omega) d \vec u d\vec \Omega \ket{0}
 \ee
 \be\label{odd}
 \ket{\psi_{odd}} = \sum_{n = 1}^{\infty} \prod_{j = 1}^{2n-1} \int  : \phi(u_{j}, \Omega_{j}): g_{2n-1}(\vec u, \vec \Omega) d \vec u d\vec \Omega \ket{0}.
 \ee
\end{subequations}
The main reason why these cases are special is because we cannot determine the value of $\sin\q_j$ (defined in eq.\eqref{thetadefn}) in the same way as we did in section \ref{sinthetagen}. This is because that method requires the presence of both even and odd $n$. We demonstrate below how that can be extended to include these special cases. 

The multiple flavoured case works in a similar way as to that of the single flavour case and therefore we shall only demonstrate the former.

\subsection{All even}
Consider a state of the form in eq.\eqref{even}. In this case, the first non-zero $g_n = g_2$. Therefore we first measure the value of $\braket{\psi|U_2^\dagg P_0 U_2|\psi}$, which allows us to determine the value of $g_2$ up to a phase. However, since the overall phase of the state $\ket{\psi}$ is meaningless, we can fix the phase of $g_2$ to be $\q_2 = 0$. 

This completely fixes the value of $g_2$. Using this it is simple to determine the value of $\cos\q_{2j}$ for all $j$. This can be performed in a similar manner as in eq.\eqref{costheta-n}, where we now have $U_2$ instead of $U_1$, i.e, measure $\braket{\psi|U_2^\dagg U_{2j}^\dagg P_0 U_{2j} U_2|\psi}$ at $\mathcal{O}(f^2)$, which gives 
\begin{eqn}
 &\cos\q_{2j} = \frac{ \braket{\psi|U_{2j}^\dagg U_2^\dagg  P_0 U_2 U_{2j}|\psi} - \braket{\psi|U_2^\dagg P_0 U_2|\psi} - \braket{\psi|U_{2j}^\dagg P_0 U_{2j}|\psi}}{2 \sqrt{\braket{\psi|U_2^\dagg P_0 U_2|\psi}} \sqrt{\braket{\psi|U_{2j}^\dagg P_0 U_{2j}|\psi}} } \,.
 \end{eqn}

 We have therefore determined the value of $\cos\q_{2j}$ for all $j$. In this measurement we also encounter the value of $\braket{\psi|U_{2j}^\dagg P_0 U_{2j}|\psi}$ which allows us to measure the value of $g_{2j+2}$ after knowing the value of $g_{2j}$.
 
 Thus the only thing that we are now left with is to find the value of $\sin\q_{2j}$, where we only need to fix its sign. This is something that can be done in two steps. We first need the value of any $\sin\q_{2j}$ other than $j =1$ (which is already fixed to be 0). By trial and error we can easily find some $2j_0$ for which $\sin\q_{2j_0} \neq 0$ (since we can measure $\cos\q_{2j_0}$). Then using the procedure as described around eq.\eqref{sinn0}, we fix the value of $\sin\q_{2j_0}$ and then go on to fix the other signs using an analogous equation to eq.\eqref{allsigns}. 
 
We explain how to carry this out for $j_0 = 2$ but in case $\sin\q_{4} = 0$, we can easily extend this for any other $j_0$. For $j_0 = 2$, we just need to measure $\braket{\psi|U_2^\dagg P_0 U_2|\psi}$ to $\mathcal{O}(f^3)$. This fixes the sign of $\sin\q_4$, although it is not enough to determine the value of $g_4$. However, since we can measure $\braket{\psi|U_4^\dagg P_0 U_4|\psi}$, we can determine $g_4$. 
 
 Once we have fixed $\sin\q_4$, we can measure $\braket{\psi|U_4^\dagg U_{2j}^\dagg P_0 U_{2j} U_4|\psi}$ at $\mathcal{O}(f^2)$ in order to determine the value of $\sin\q_{2j}$ completely. Therefore we have demonstrated how one can fix the functions $g_{2j}$ including the phase factors $\q_{2j}$ completely, in a sieve-like manner.

\subsection{All odd}
This is very similar to the previous case, therefore we only list the important steps.
\begin{enumerate}
 \item We first fix the value of $g_1$ by measuring $\braket{\psi|U_1^\dagg P_0 U_1|\psi}$ and then using the fact that the overall phase of the state $\ket{\psi}$ is meaningless, we fix $\q_1 = 0$. 
 \item Then we measure the value of $\braket{\psi|U_1^\dagg U_{2j-1}^\dagg P_0 U_{2j-1} U_1|\psi}$ at $\mathcal{O}(f^2)$ to fix the value of $\cos\q_{2j-1}$. 
 \item Measuring $\braket{\psi|U_{2j-1}^\dagg P_0 U_{2j-1}|\psi}$ at $\mathcal{O}(f^2)$ yields the value of $g_{2j-1}$ up to the phase factor $e^{i \q_{2j-1}}$. This step has to be performed in a sieve-like procedure; the value of $\cos\q_{2j-1}$ is  fixed from the previous step and therefore we are left with fixing the sign of $\sin\q_{2j-1}$ only. 
 \item In order to fix the signs of all $\sin\q_{2j-1}$ we first fix the value of any particular $\sin\q_{2j-1} \neq 0$ and then performing a measurement of the kind in eq.\eqref{allsigns}, allows to fix everything else. For the case when $\sin\q_{3} \neq 0$ (which is an example, that can be extended to any $2j-1$) we have to measure $\braket{\psi|U_2^\dagg U_1^\dagg P_0 U_1 U_2|\psi}$ at $\mathcal{O}(f^3)$ in order to fix the sign of $\sin\q_3$. 
 \item By measuring $\braket{\psi|U_3^\dagg U_{2j - 1}^\dagg P_0 U_{2j - 1} U_3|\psi}$ we can fix the signs of all $\sin\q_{2j -1}$. 
\end{enumerate}

\end{appendix}
\printbibliography 
\end{document}